\documentclass[conference,10pt]{IEEEtran}
\IEEEoverridecommandlockouts
\usepackage[colorlinks=false]{hyperref}
\usepackage{amsmath,amssymb,amsfonts}
\usepackage{adjustbox}
\usepackage{algorithmic}
\usepackage{threeparttable}
\usepackage{array}
\usepackage{fontawesome}
\usepackage{graphicx}
\usepackage{multirow}
\usepackage{textcomp}
\usepackage{enumitem}
\usepackage[english]{babel}
\usepackage[utf8]{inputenc}
\usepackage[nolist]{acronym}
\usepackage{amsmath}											
\usepackage{amsfonts}											
\usepackage{blindtext} 											
\usepackage{booktabs}											
\usepackage{fancyvrb}
\usepackage{ifthen}	
\usepackage{wrapfig}											
\usepackage{filecontents} 										
\usepackage[T1]{fontenc} 
\usepackage{graphicx}
\usepackage[dvipsnames,table]{xcolor}
\usepackage{ifthen}												
\usepackage{listings}											
\usepackage{lipsum}
\usepackage{multirow}
\usepackage{pgfplots}
\usepackage{siunitx}
\usepackage{ragged2e}						
\usepackage{tabularx}
\usepackage{url}
\usepackage{tikz}
\usepackage{circuitikz}
\usepackage{comment}
\usepackage{footnote}
\usepackage{float}
\usepackage{balance}
\usepackage{setspace}
\usepackage{makecell, tabularx}
\newtheorem{theorem}{Theorem}[section] 
\newtheorem{lemma}[theorem]{Lemma}    

\usetikzlibrary{calc,trees,positioning,arrows,chains,shapes.geometric,%
	decorations.pathreplacing,decorations.pathmorphing,shapes,%
	matrix,shapes.symbols}

\usepackage[font=footnotesize,labelfont=bf]{subcaption}		
\usepackage{caption}
\usepackage[ruled,linesnumbered]{algorithm2e}
\SetAlCapNameFnt{\small}
\SetAlCapFnt{\small}
\SetAlgoNlRelativeSize{0}
\RestyleAlgo{ruled}
\SetKwComment{Comment}{/* }{ */}

\usepackage[style=ieee, backend=biber, natbib=true, maxbibnames=1, minbibnames=1]{biblatex}
\usepackage{fancyhdr}

\fancypagestyle{firstpage}{%
	\fancyhf{}%
	\fancyhead[C]{\normalsize To appear at the 37th IEEE International Conference on Microelectronics  (ICM), December 14-17, 2025, Cairo, Egypt}
	
	\fancyfoot[C]{\footnotesize \copyright2025 IEEE.  Personal use of this material is permitted.  Permission from IEEE must be obtained for all other uses, in any current or future media, including reprinting/republishing this material for advertising or promotional purposes, creating new collective works, for resale or redistribution to servers or lists, or reuse of any copyrighted component of this work in other works.}
}

\lstdefinelanguage{Verilog}{morekeywords={accept_on,alias,always,always_comb,always_ff,always_latch,and,assert,assign,assume,automatic,before,begin,bind,bins,binsof,bit,break,buf,bufif0,bufif1,byte,case,casex,casez,cell,chandle,checker,class,clocking,cmos,config,const,constraint,context,continue,cover,covergroup,coverpoint,cross,deassign,default,defparam,design,disable,dist,do,edge,else,end,endcase,endchecker,endclass,endclocking,endconfig,endfunction,endgenerate,endgroup,endinterface,endmodule,endpackage,endprimitive,endprogram,endproperty,endspecify,endsequence,endtable,endtask,enum,event,eventually,expect,export,extends,extern,final,first_match,for,force,foreach,forever,fork,forkjoin,function,generate,genvar,global,highz0,highz1,if,iff,ifnone,ignore_bins,illegal_bins,implements,implies,import,incdir,include,initial,inout,input,inside,instance,int,integer,interconnect,interface,intersect,join,join_any,join_none,large,let,liblist,library,local,localparam,logic,longint,macromodule,matches,medium,modport,module,nand,negedge,nettype,new,nexttime,nmos,nor,noshowcancelled,not,notif0,notif1,null,or,output,package,packed,parameter,pmos,posedge,primitive,priority,program,property,protected,pull0,pull1,pulldown,pullup,pulsestyle_ondetect,pulsestyle_onevent,pure,rand,randc,randcase,randsequence,rcmos,real,realtime,ref,reg,reject_on,release,repeat,restrict,return,rnmos,rpmos,rtran,rtranif0,rtranif1,s_always,s_eventually,s_nexttime,s_until,s_until_with,scalared,sequence,shortint,shortreal,showcancelled,signed,small,soft,solve,specify,specparam,static,string,strong,strong0,strong1,struct,super,supply0,supply1,sync_accept_on,sync_reject_on,table,tagged,task,this,throughout,time,timeprecision,timeunit,tran,tranif0,tranif1,tri,tri0,tri1,triand,trior,trireg,type,typedef,union,unique,unique0,unsigned,until,until_with,untyped,use,uwire,var,vectored,virtual,void,wait,wait_order,wand,weak,weak0,weak1,while,wildcard,wire,with,within,wor,xnor,xor,`uvm_create, `uvm_rand_send_with},morecomment=[l]{//}}

\newboolean{blindreview}
\setboolean{blindreview}{false}
\DeclareRobustCommand{\IEEEauthorrefmark}[1]{\smash{\textsuperscript{\footnotesize #1}}}

\usepackage[left=1.57cm,right=1.57cm,top=2cm,bottom=3cm,letterpaper]{geometry}

\begin{document}	
\begin{acronym}[] 
	\acro{BMC}[BMC]{Bounded Model Checker}
	\acro{CBMC}[CBMC]{C Bounded Model Checker}
	\acro{CEX}[CEX]{Counterexample}
	\acrodefplural{CEX}[CEXs]{counterexamples}
	\acro{COI}[COI]{cone of influence}
	\acro{EDA}[EDA]{Electronic Design Automation}
	\acro{FPU}[FPU]{Floating-Point Unit}
	\acrodefplural{FPUs}[FPUs]{Floating-Point Units}
	\acro{HLS}[HLS]{High-Level Synthesis}
	\acro{HITL}[HITL]{Human‑in‑the‑Loop}
	\acro{LLM}[LLM]{large language model}
	\acrodefplural{LLM}[LLMs]{large language models}
	\acro{MAS}[MAS]{multi‑agent system}
	\acro{NaN}[NaN]{Not a Number}
	\acro{PDD}[PDD]{Property-Driven Hardware Design}
	\acro{RTL}[RTL]{Register Transfer Level}
	\acro{SAT}[SAT]{Boolean Satisfiability}
	\acro{SMT}[SMT]{Satisfiability Modulo Theories}
	\acro{SLEC}[SLEC]{Sequential Logic Equivalence Checking}
	\acrodefplural{SVA}[SVAs]{SystemVerilog Assertions}
	\acro{vPlan}[vPlan]{verification Plan}
	\acro{V2C}[V2C]{Verilog-to-C}
\end{acronym}


\title{Formal that "Floats" High: \\
	Formal Verification of Floating Point Arithmetic

}

\ifthenelse{\boolean{blindreview}}
{}{
	\author{
		\IEEEauthorblockN{
			Hansa Mohanty\IEEEauthorrefmark{1},
			Vaisakh Naduvodi Viswambharan\IEEEauthorrefmark{2},
			Deepak Narayan Gadde\IEEEauthorrefmark{2}
			}
		\IEEEauthorblockA{
			\IEEEauthorrefmark{1}Infineon Technologies Semiconductor India Private Limited, India\\
			\IEEEauthorrefmark{2}Infineon Technologies Dresden AG \& Co. KG, Germany
			}
		}
}

\maketitle

\thispagestyle{firstpage}

\begin{abstract}

Formal verification of floating-point arithmetic remains challenging due to non-linear arithmetic behavior and the tight coupling between control and datapath logic. Existing approaches often rely on high-level C models for equivalence checking against \ac{RTL} designs, but this introduces abstraction gaps, translation overhead, and limits scalability at the \ac{RTL} level. To address these challenges, this paper presents a scalable methodology for verifying floating-point arithmetic using direct \ac{RTL}-to-\ac{RTL} model checking against a golden reference model. The approach adopts a divide-and-conquer strategy that decomposes verification into modular stages, each captured by helper assertions and lemmas that collectively prove a main correctness theorem. \ac{CEX}-guided refinement is used to iteratively localize and resolve implementation defects, while targeted fault injection validates the robustness of the verification process against precision-critical datapath errors. To assess scalability and practicality, the methodology is extended with agentic AI-based formal property generation, integrating \ac{LLM}-driven automation with \ac{HITL} refinement. Coverage analysis evaluates the effectiveness of the approach by comparing handwritten and AI-generated properties in both \ac{RTL}-to-\ac{RTL} model checking and standalone \ac{RTL} verification settings. Results show that direct \ac{RTL}-to-\ac{RTL} model checking achieves higher coverage efficiency and requires fewer assertions than standalone verification, especially when combined with AI-generated properties refined through \ac{HITL} guidance.

\end{abstract}


\section{Introduction} \label{sec:introduction}
Ensuring the correctness of hardware designs is critical in modern computing, as even minor flaws can cause catastrophic failures in safety-critical and high-performance systems. Formal verification provides mathematically rigorous guarantees of correctness by exhaustively analyzing all possible behaviors of a design \cite{10.1145/307988.307989,siemens2024formal}. Unlike simulation, which tests only a subset of inputs, formal verification provides exhaustive state-space exploration \cite{bsi_formal_methods,hsieh2024formal,ho2010post_silicon_debug}, enabling early detection of logical inconsistencies and corner-case bugs. 

Verifying floating-point arithmetic is particularly challenging due to the tight interplay between discrete control logic and arithmetic datapaths, together with strict IEEE-754 compliance requirements \cite{8766229}. \acp{FPU} include complex operations such as exponent alignment, normalization, and rounding, each of which introduces edge cases such as denormalized operands, overflow, underflow, and precision loss. The interaction between control logic (e.g., exception handling) and arithmetic datapaths makes exhaustive \ac{RTL}-level verification difficult, placing floating-point verification among the most demanding formal verification tasks.

This paper presents a scalable formal verification methodology for arithmetic datapaths, demonstrated on a floating-point adder. The approach applies direct \ac{RTL}-to-\ac{RTL} model checking to compare an implementation against a cycle-accurate golden reference model. Using a divide-and-conquer strategy, the design is decomposed hierarchically into modular verification stages, each verified using property-driven checking. \acp{CEX} from the formal tool guide iterative refinement until full equivalence is proven.

The main contributions of this paper are as follows:

\begin{itemize}
	\item \textit{Direct \ac{RTL}-to-\ac{RTL} model checking:} Verifies the \ac{RTL} implementation against a golden reference model without relying on high-level C models (Section~\ref{subsec:direct_rtl_to_rtl_model_checking}).
	
	\item \textit{Property-driven verification:} Applies a divide-and-conquer strategy that partitions the design into modular stages and uses \ac{CEX}-guided refinement to ensure correctness (Section~\ref{subsec:propertydriven_hierarchicalverification}).
	
	\item \textit{Bug detection and formal proof:} Identifies and resolves implementation defects and proves correctness against the reference model (Section~\ref{subsec:bugdetection_formalproof}).
	
	\item \textit{Fault injection:} Evaluates design resilience and validates property effectiveness through systematic fault injection (Section~\ref{subsec:fault_injection}).
	
	\item \textit{Coverage analysis:} Compares \ac{RTL}-to-\ac{RTL} model checking and standalone \ac{RTL} verification using handwritten and AI-generated formal properties (Section~\ref{subsec:coverage_analysis}).
\end{itemize}

The remainder of this paper is organized as follows:  Section~\ref{sec:background} reviews related work, Section~\ref{sec:methodology} details the methodology, Section~\ref{sec:results} presents evaluation results, and  Section~\ref{sec:conclusion} concludes with key takeaways.

\section{Related Work} \label{sec:background}
Formal verification of both control-path and datapath circuits has been extensively explored using model checking, theorem proving, equivalence checking, and property-driven techniques. However, a significant portion of prior work relies on C-based modeling or translation flows to enable verification, introducing an abstraction gap between the specification and the hardware implementation. While effective at higher levels of abstraction, these approaches face limitations when applied directly to datapath-intensive \ac{RTL} designs due to scalability constraints, semantic mismatches between C and hardware execution, and increased manual effort for model maintenance.

A significant body of work relies on C-based verification toolchains such as \ac{CBMC}. Tools like CREST \cite{tiemeyer2019cresthardwareformalverification} translate \ac{RTL} into C and use \ac{SAT} or \ac{SMT} solving to verify floating-point correctness. However, these methods depend on high-level ANSI-C specifications and introduce semantic translation overhead. Other automated toolchains \cite{7308670,8551480} also translate \ac{V2C} or intermediate forms for \ac{BMC} and symbolic execution, but this multi-language flow reduces efficiency for datapath-heavy \ac{RTL} verification.

Equivalence checking approaches such as DEEQ \cite{9806218} compare C/C++ specifications with \ac{RTL} generated by \ac{HLS}, but their applicability is restricted to \ac{HLS}-based workflows. Property-driven methods \cite{8759950,9489148} improve modular verification but still rely on abstract SystemC models, introducing modeling effort and limiting scalability for datapath-intensive designs. Recent \ac{RTL}-to-software equivalence checking for floating-point units \cite{morini2024dpas} reduces abstraction mismatch but depends on proprietary C++ models and closed verification flows. Commercial tools such as Siemens \ac{SLEC} \cite{pouarz2024sle} support bit-accurate floating-point validation but require extensive setup effort and high-quality software reference models.

In summary, prior work advances equivalence checking and model abstraction but relies heavily on C/C++ reference models, translation flows, or intermediate representations, which limit applicability to direct \ac{RTL}-level datapath verification. To address this gap, the proposed work introduces a direct \ac{RTL}-to-\ac{RTL} formal verification methodology that removes dependence on software abstractions and HLS-generated models, enabling precise and scalable verification of arithmetic datapaths. Table~\ref{tab:related_work} provides a structured comparison of the referenced methodologies.

\begin{table*}[htbp!]
	\caption{Comparison of Related Works}
	\label{tab:related_work}
	\centering
	\scriptsize 
	\begin{tabular}{|p{1cm}|p{0.4cm}|p{3.6cm}|p{2.4cm}|p{3.6cm}|p{4cm}|}
		\hline
		\textbf{Work} & \textbf{Year} & \textbf{Methodology} & \textbf{DUV} & \textbf{Tools} & \textbf{Dependencies} \\ \hline
		
		\cite{7308670} & 2015 & Verilog-to-C translation & Serial adder, Divider & V2C, \ac{CBMC}, Path-Symex, Astrée &  Formal software analyzers\\ \hline
		
		\cite{pouarz2024sle} & 2017 & C-to-\ac{RTL} equivalence checking & Floating-point units & Siemens SLEC System & High-quality reference models in C++ \\ \hline
		
		\cite{8551480} & 2018 & Verilog-to-C-to-Verilog translation & Adder, 64-bit Processor & V2C, \ac{CBMC}, Bambu & Syntax error rectification for V2C output \\ \hline
		
		\cite{tiemeyer2019cresthardwareformalverification} & 2019 & C-to-Verilog translation & Floating-point units & CREST - adaptation of \ac{CBMC} & High-quality ANSI-C specification \\ \hline
		
		\cite{8759950} & 2020 & Property-Driven Hardware Design & FPI bus, Wishbone bus & DeSCAM, OneSpin 360 DV-Certify & SystemC-PPA for abstract modeling \\ \hline
		
		\cite{9489148} & 2021 & Property-Driven Hardware Design & FIR filter & DeSCAM, QuestaSim & SystemC-PPA for abstract modeling \\ \hline
		
		\cite{9806218} & 2022 & C-to-\ac{RTL} equivalence checking & MatrixAdd, Dfadd & Vivado HLS, pyVerilog, Klee &  FSMD modeling\\ \hline
		
		\cite{morini2024dpas} & 2024 & \ac{RTL}-to-C++ equivalence checking & DPAS unit & - & Proprietary iFP-based C++ models \\ \hline
		
		This work & 2025 & \ac{RTL}-to-\ac{RTL} model checking & Floating-point adder & Cadence Jasper & High-quality RTL reference model \\ \hline
		
	\end{tabular}
	
	\noindent
	\begin{minipage}{\textwidth}
		\vspace{1em}
		\scriptsize 
		Notes: Design Under Verification (DUV), Flexible Peripheral Interconnect (FPI), Path Predicate Abstraction (PPA), Finite Impulse Response (FIR), Finite State Machines with Datapath (FSMD),  Dot Product Accumulate Systolic (DPAS)
	\end{minipage}

\end{table*}

\section{Methodology} \label{sec:methodology}
This section outlines the formal verification approach for \acp{FPU} using direct \ac{RTL}-to-\ac{RTL} model checking. It is organized into four subsections: an overview of the proposed methodology, a description of the floating-point adder design, a detailed explanation of the verification process, and an agentic AI-based system for automated formal property generation.

\begin{figure}[ht!]
	\centering
	\includegraphics[width=0.9\linewidth]{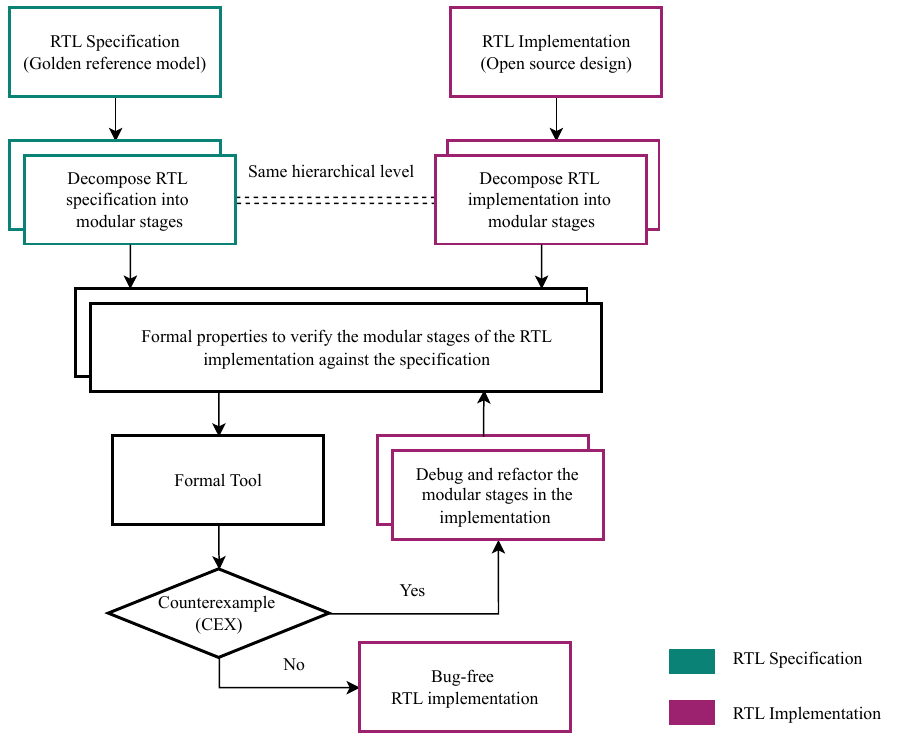}
	\caption{Flowchart of the proposed RTL-to-RTL model checker.}
	\label{fig:block_diagram}
\end{figure}

\subsection{Overview}
\label{subsec:direct_rtl_to_rtl_model_checking}

Formal verification of complex datapath designs requires a systematic strategy to ensure functional correctness at the \ac{RTL} level. As shown in Fig.~\ref{fig:block_diagram}, the proposed methodology applies direct \ac{RTL}-to-\ac{RTL} model checking to verify the implementation against a golden reference model. Hierarchical decomposition partitions both designs into modular stages, enabling stage-wise verification that reduces the \ac{COI}, lowers proof complexity, and improves convergence during model checking.

Formal properties are defined for each stage and exhaustively evaluated using an industry‑standard property checker. When a mismatch is detected, the failing stage is isolated using the generated \ac{CEX}, and the implementation or property set is refined to resolve the discrepancy. This refinement cycle continues until all properties are proven, establishing functional equivalence between the two designs.

\subsection{Floating-Point Adder Design}

A floating-point number is expressed as a triple $(s, e, m)$, where $s$ is the sign bit, $e$ is the exponent, and $m$ is the mantissa (or significand). 
Its numerical value is given by:

\begin{equation}\scriptsize
	(-1)^s \times 2^{e - bias} \times (1.m)
\end{equation}

Here, $bias$ is the format-dependent exponent bias, and the mantissa has an implicit leading 1 in normalized representations, ensuring that $1.m$ lies within the range $[1, 2)$. 
Floating-point addition for two normalized operands \(f1 = (s1, e1, m1)\) and \(f2 = (s2, e2, m2)\) follows the computational steps in Equations~\ref{eq2}–\ref{eq8}.

\begingroup
\scriptsize
\begin{equation} \label{eq2}
	expdiff =  |e_1 -e_2|,
\end{equation}
\begin{equation} \label{eq3}
	bigman =
	\begin{cases} 
		1.m_1, & \text{if } e_1 \leq e_2, \\
		1.m_2, & \text{otherwise.}
	\end{cases}
\end{equation}
\begin{equation} \label{eq4}
	smallman =
	\begin{cases} 
		1.m_2, & \text{if } e_1 \leq e_2, \\
		1.m_1, & \text{otherwise.}
	\end{cases}          
\end{equation}
\begin{equation} \label{eq5}
	algman = shift\_right(smallman, expdiff),
\end{equation}
\begin{equation} \label{eq6}
	addman = bigman+algman,
\end{equation}
\begin{equation} \label{eq7}
	m = round(normalize(addman)),
\end{equation}
\begin{equation} \label{eq8}
	result = (s,e,m)
\end{equation}
\endgroup

The floating-point adder used in this study is implemented in SystemVerilog and sourced from an open-source repository \cite{pulp-fpu}. The design is fully combinational, which simplifies formal analysis by avoiding sequential state exploration. A golden reference model, also in SystemVerilog, provides a precise, bit-accurate IEEE-754 compliant implementation. Although it mirrors the computation steps of the adder, it prioritizes correctness over hardware optimization, making it suitable for equivalence checking.

\subsection{Implementation of Methodology}

This section details the implementation of the proposed formal verification methodology for the floating-point adder. The goal is to prove that the \ac{RTL} implementation conforms exactly to a golden reference model using direct \ac{RTL}-to-\ac{RTL} model checking. The methodology employs hierarchical decomposition and structured correctness proof using a main theorem supported by intermediate lemmas and stage-level assertions.

\begin{figure}[ht!]
	\centering
	\includegraphics[width=0.9\linewidth]{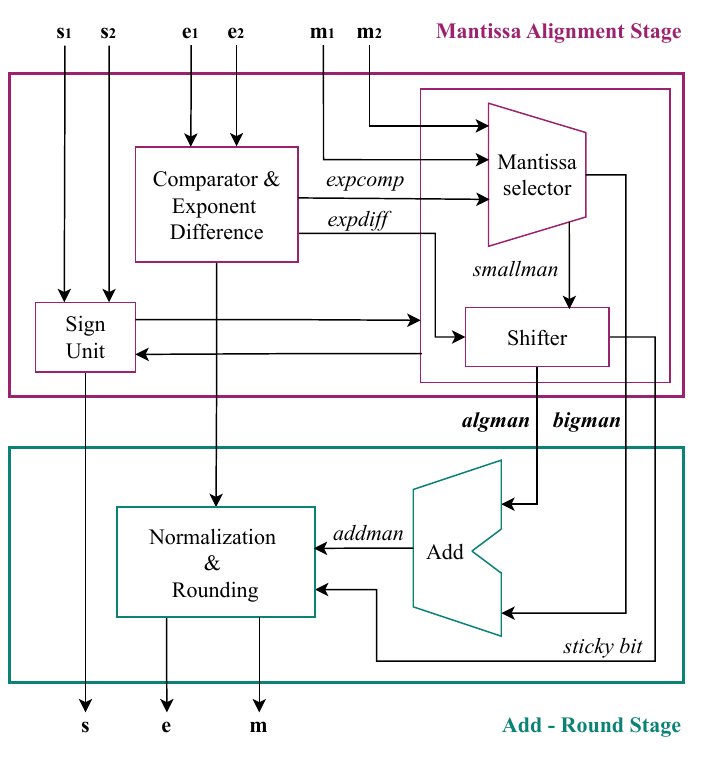}
	\caption{Hierarchical decomposition of the floating-point adder into modular stages: Mantissa Alignment Stage and Add-Round Stage.}
	\label{fig:adder}
\end{figure}

In this context, a theorem represents a high-level correctness requirement, while lemmas capture intermediate correctness guarantees necessary to prove the theorem. This decomposition simplifies reasoning by isolating independent computational stages, improving scalability and proof convergence. The primary correctness condition for the floating-point adder is captured in the following theorem:

\begin{theorem}[Result Equivalence]
	\label{theorem:result}
	For all valid normalized floating-point input pairs \(f_1\) and \(f_2\), the final rounded floating-point result produced by the implementation must be functionally equivalent to that of the specification:
	\begin{align*}
			spec.result = impl.result
	\end{align*}
\end{theorem}

To verify Theorem~\ref{theorem:result}, the floating-point adder is decomposed into two modular stages: the Mantissa Alignment Stage and the Add-Round Stage, as described in \cite{beers2001applications} and shown in Fig.~\ref{fig:adder}. 
This hierarchical approach reduces complexity and facilitates debugging of intermediate outputs.

The Mantissa Alignment Stage aligns the mantissas of the two input floating-point numbers by right-shifting the smaller mantissa based on the absolute exponent difference (\(expdiff\)). Equations~\ref{eq2} to \ref{eq5} define the key computations in this stage, such as determining the exponent difference, selecting the larger and smaller mantissas, and performing alignment. These computations are captured in the following lemma:

\begin{lemma}[Mantissa Alignment Equivalence] 
	\label{lemma:alignment}
	Given two normalized floating-point inputs \(f_1\) and \(f_2\), the aligned mantissa (\(algman\)) and the larger mantissa (\(bigman\)) produced by the implementation must match those of the specification:
	\begin{align*}
		& (spec.algman = impl.algman) \\
		& \land (spec.bigman = impl.bigman)
	\end{align*}
\end{lemma}

Lemma~\ref{lemma:alignment} is verified using stage-level supporting assertions, such as the one shown in Listing~\ref{MantissaAlignmentAssertion}, which ensures that the implementation correctly performs exponent comparison and mantissa alignment.

\begin{lstlisting}[
	language=Verilog,
	caption=Assertion to verify Lemma~\ref{lemma:alignment},
	label={MantissaAlignmentAssertion},
	]
	property mantissa_align_equivalence;
	(impl.s1==spec.s1) && (impl.s2==spec.s2) && 
	// Signs of input operands are same
	(impl.e1==spec.e1) && (impl.e2==spec.e2) && 
	// Exponents of input operands are equal
	(impl.m1==spec.m1) && (impl.m2==spec.m2) 
	// Mantissas of input operands are equal
	|->
	(impl.algman==spec.algman) && 
	// Aligned mantissas must be equal
	(impl.bigman==spec.bigman);
	// Larger mantissas must be equal
	endproperty
	
	ap_mantissa_align_equivalence : 
	assert property(mantissa_align_equivalence);
	
\end{lstlisting}

The Add-Round Stage performs addition of the aligned mantissas, followed by normalization and rounding. 
Equations~\ref{eq6} to \ref{eq8} detail the computations in this stage, which are formalized in the following lemma:

\begin{lemma}[Add-Round Equivalence] 
	\label{lemma:rounding}
	Given that the aligned mantissa (\(algman\)) and the larger mantissa (\(bigman\)) are identical between the implementation and specification in the Mantissa Alignment Stage, the final result of the Add-Round Stage, computed through addition, normalization, and rounding, must also match between the implementation and the specification.
\end{lemma}

The correctness condition in Lemma~\ref{lemma:rounding} is verified through helper assertions in Listing~\ref{AddRoundAssertion}. These assertions confirm equivalence between the implementation and specification results.

In \cite{beers2001applications}, the equivalence check for the Add-Round Stage required additional good properties to constrain the reference model's inputs, addressing architectural differences between the reference model and the implementation. These properties refined the reference model to account for implementation-specific behavior, enabling equivalence verification at the Add-Round Stage output. In contrast, this work eliminates the need for such refinements, as the reference specification and implementation are architecturally aligned. As a result, Lemma~\ref{lemma:rounding} directly verifies equivalence between the specification and implementation without additional constraints.

\begin{lstlisting}[
	language=Verilog, 
	caption=Assertion to verify Lemma~\ref{lemma:rounding}, 
	label={AddRoundAssertion},
	]
	property add_round_equivalence;
	(impl.s1 == spec.s1) && (impl.s2 == spec.s2) && 
	// Signs of input operands are same
	(impl.algman == spec.algman) && 
	// Aligned mantissas are equal
	(impl.bigman == spec.bigman)
	// Larger mantissas are equal
	|-> 
	(impl.s == spec.s) && 
	// Sign of the results must be equal
	(impl.e == spec.e) &&
	// Exponent of the results must be equal
	(impl.m == spec.m));
	// Mantissa of the results must be equal
	endproperty
	
	property exp_inputs_are_equal;
	1 |-> ((impl.e1 == spec.e1) && 
	(impl.e2 == spec.e2))
	// Exponents of input operands are equal
	endproperty
	
	ap_add_round_equivalence: 
	assert property(add_round_equivalence);
	
	cp_exp_inputs_are_equal:
	assume property(exp_inputs_are_equal);
\end{lstlisting}

Theorem~\ref{theorem:result} is established by proving Lemma~\ref{lemma:alignment} and Lemma~\ref{lemma:rounding}, confirming correctness of both computational stages. This structured approach simplifies verification, enables targeted debugging, improves proof convergence, and ensures high confidence in the correctness of the \ac{RTL} implementation.

\subsection{Agentic AI-Based Formal Property Generation}

\begin{figure}[ht!]
	\centering
	\includegraphics[width=0.9\linewidth]{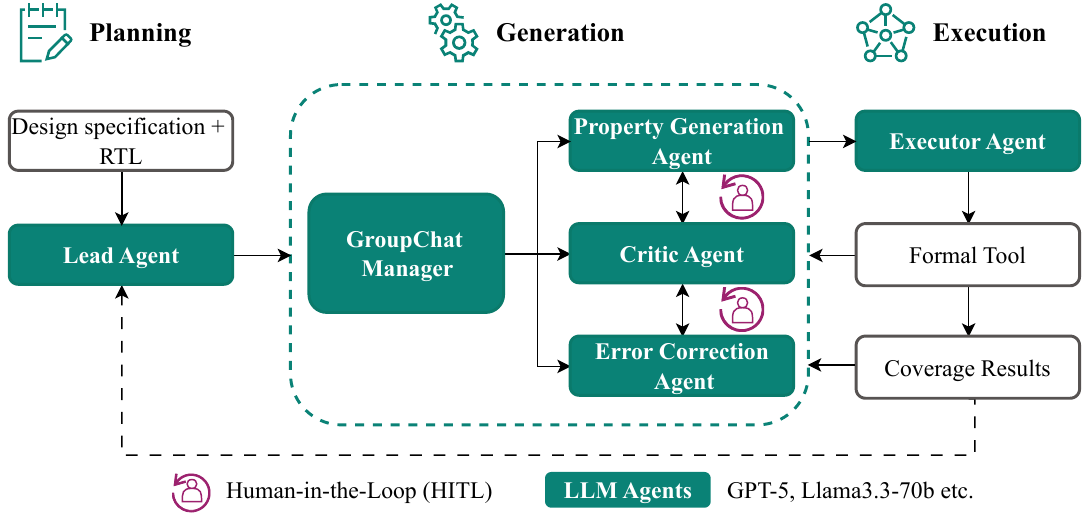}
	\caption{Agentic AI-based formal verification workflow \cite{11218681}}
	\label{fig:agent_orchestration}
\end{figure}

A \ac{MAS} was employed to generate \acp{SVA} from natural language specifications, with iterative refinement under \ac{HITL} supervision \cite{11218681}. As shown in Fig.~\ref{fig:agent_orchestration}, the workflow consists of three coordinated stages: \textit{Planning}, \textit{Generation}, and \textit{Execution}.

In the \textit{planning stage}, a \textit{Lead Agent} interprets natural-language design intent and \ac{RTL} behavior to produce a structured \ac{vPlan} that organizes properties into functional categories and defines coverage goals.

In the \textit{generation stage}, the \textit{Group Chat Manager} coordinates collaboration among three specialized agents:
\begin{itemize}
	\item \textit{Property Generation Agent} – uses \acp{LLM} (e.g., GPT-5) to translate \ac{vPlan} entries into IEEE-754 compliant \acp{SVA},
	\item \textit{Critic Agent} – evaluates logical correctness, consistency with design intent, and coverage relevance,
	\item \textit{Error Correction Agent} – resolves syntax and semantic issues identified during review.
\end{itemize}
If convergence of property correctness is not achieved within a bounded number of refinement cycles, \ac{HITL} guidance is introduced to clarify design intent and resolve ambiguities.

In the \textit{execution stage}, the \textit{Executor Agent} drives formal verification by submitting generated properties to a property checker, analyzing \acp{CEX}, and monitoring coverage. Coverage gaps trigger targeted regeneration of new properties, maintaining alignment with verification objectives.

This event-driven multi-agent workflow combines automated property generation with semantic validation and iterative refinement, enabling verification of both \ac{RTL}-to-\ac{RTL} model checking and standalone \ac{RTL} verification.

\section{Evaluation} \label{sec:results}
The \ac{RTL}-to-\ac{RTL} verification experiments were conducted using the Cadence Jasper Formal Verification Platform \cite{cadence_jasper}. The objective was to prove Lemma~\ref{lemma:alignment} and Lemma~\ref{lemma:rounding}, thereby establishing functional equivalence between the implementation and the golden reference model.

\subsection{Mantissa Alignment Stage}
\label{subsec:bugdetection_formalproof}

During verification, the \(mantissa\_align\_equivalence\) property produced \acp{CEX}, indicating discrepancies in \(algman\) and \(bigman\) generation between the implementation and reference model. Analysis revealed two faults in the operand preparation logic:

\begin{itemize}
	\item \textit{Incorrect operand selection:} When input exponents were equal, the implementation failed to compare mantissas to select the larger operand, leading to misalignment.
	\item \textit{Faulty inversion logic:} Mantissa inversion during subtraction triggered incorrectly in certain cases, propagating bit errors.
\end{itemize}

These faults were corrected by enforcing consistent mantissa comparison for operand selection and refining inversion conditions based on sign and magnitude. After refinement, the implementation satisfied the \(mantissa\_align\_equivalence\) property with no further \acp{CEX}.

\subsection{Add-Round Stage}
\label{subsec:propertydriven_hierarchicalverification}

Initial verification of the Add-Round Stage confirmed mantissa equivalence but revealed mismatches in result sign and exponent. Two key faults were identified:

\begin{itemize}
	\item \textit{Unconstrained exponent behavior:} Allowed invalid exponent propagation, corrupting sign computation.
	\item \textit{Exponent mismatch between instances:} Enabled operand interchange between the reference model and implementation, producing incorrect equivalence.
\end{itemize}

The assertion was refined by constraining exponent difference (\(expdiff\)), enforcing consistent operand ordering, and matching input exponent pairs (\(e1\), \(e2\)) between implementation and reference. After refinement, the properties were proven without \acp{CEX}, confirming functional equivalence.

\subsection{Fault Injection}
\label{subsec:fault_injection}

Fault injection was employed to evaluate the robustness of the verification process and the ability of formal properties to detect precision-critical defects. Controlled faults were systematically introduced into the \ac{RTL} implementation to emulate common datapath design errors and assess their impact relative to the golden reference model.

In the Mantissa Alignment Stage, four faults were injected:
\begin{enumerate}
	\item \textit{Sticky-bit distortion:} An incorrect shift offset in sticky-bit computation caused loss of alignment precision.
	\item \textit{Operand extension misalignment:} Missing zero padding during mantissa extension produced incorrect binary alignment.
	\item \textit{Faulty operand selection:} A reversed comparison condition led to incorrect identification of the larger operand.
	\item \textit{Inversion swap in subtraction:} Two’s complement inversion was incorrectly applied to the wrong operand, altering subtraction behavior.
\end{enumerate}

These faults affected operand preparation and alignment accuracy, propagating incorrect values into subsequent computations and resulting in detectable mismatches with the reference model.

In the Add‑Round Stage, four faults were injected:
\begin{enumerate}
	\item \textit{Carry-in manipulation:} Missing or extra carry bits corrupted mantissa addition.
	\item \textit{Normalization shift error:} Faulty leading-one detection combined with missing exponent correction caused mantissa and exponent misalignment.
	\item \textit{Shift distortion:} Shifting the extended sum by an incorrect offset misaligned the normalized mantissa.
	\item \textit{Rounding rule violation:} Modifying the rounding comparison logic violated the round-to-nearest-even rule.
\end{enumerate}

All injected faults introduced divergences at the \ac{RTL} level and were successfully detected by the formal properties, demonstrating strong fault sensitivity and verification completeness.

\subsection{Coverage Analysis}
\label{subsec:coverage_analysis}

To evaluate robustness across verification scenarios, two comparative assessments were conducted. First, \ac{RTL}-to-\ac{RTL} model checking was evaluated using both handwritten and \ac{LLM}-generated properties. Second, \ac{LLM}-generated properties were analyzed in two verification settings: with a golden reference and without a reference model. 
For each \ac{LLM}, the multi-agent workflow generated and refined an initial property set and then invoked the formal tool once to measure coverage. Coverage gaps from this run triggered a second, targeted property-generation iteration, after which the formal tool was invoked a final time. The reported proof times and coverage values correspond to this final run.

\begin{table*}[ht!]
	\centering
	\caption{Overview of the results}
	\label{tab:results}
	\setcellgapes{1.2pt}
	\scriptsize
	\renewcommand\arraystretch{1.2}
	
	\makegapedcells
	\begin{tabular}{ccccccccccccc}
		\hline
		\multirow{2}{*}{\textbf{Methodology}} &
		\multirow{2}{*}{\textbf{\begin{tabular}[c]{@{}c@{}}Mode of \\ generation\end{tabular}}} &
		\multirow{2}{*}{\textbf{\begin{tabular}[c]{@{}c@{}}HITL \\ feedback\end{tabular}}} &
		\multicolumn{3}{c}{\textbf{\#Assertions}} &
		\multirow{2}{*}{\textbf{\begin{tabular}[c]{@{}c@{}}Proof \\ Time(s)\end{tabular}}} &
		\multicolumn{3}{c}{\textbf{\#Cover Items}} &
		\multicolumn{3}{c}{\textbf{Coverage Type (\%)}} \\ \cline{4-6} \cline{8-13} 
		&
		&
		&
		\textbf{Total} &
		\textbf{Proven} &
		\textbf{Failed} &
		&
		\textbf{Total} &
		\textbf{Covered} &
		\textbf{Unreachable} &
		\textbf{Formal} &
		\textbf{Stimuli} &
		\textbf{Checker} \\ \hline
		\multirow{9}{*}{\textbf{\begin{tabular}[c]{@{}c@{}}RTL-to-RTL \\ model checking \\ with \\ formal property \\ verification\end{tabular}}} &
		Handwritten &
		- &
		2 &
		2 &
		0 &
		0.073 &
		\multirow{9}{*}{98} &
		\multirow{9}{*}{92} &
		\multirow{9}{*}{6} &
		\multirow{9}{*}{98.42} &
		\multirow{9}{*}{100} &
		\multirow{9}{*}{98.42} \\ \cline{2-7}
		& \multirow{2}{*}{Llama3.3-70b} & No  & 7  & 6  & 1 & 5.822*  &  &    &    &       &  &       \\
		&                               & Yes & 6  & 6  & 0 & 6.581   &  &    &    &       &  &       \\ \cline{2-7}
		& \multirow{2}{*}{GPT-4o}       & No  & 9  & 3  & 6 & 1.697*  &  &    &    &       &  &       \\
		&                               & Yes & 6  & 6  & 0 & 2.538   &  &    &    &       &  &       \\ \cline{2-7}
		& \multirow{2}{*}{GPT-4.1}      & No  & 8  & 4  & 4 & 3.149*  &  &    &    &       &  &       \\
		&                               & Yes & 6  & 6  & 0 & 7.949   &  &    &    &       &  &       \\ \cline{2-7}
		& \multirow{2}{*}{GPT-5}        & No  & 15 & 14 & 1 & 12.828* &  &    &    &       &  &       \\
		&                               & Yes & 3  & 3  & 0 & 1.318   &  &    &    &       &  &       \\ \hline
		\multirow{8}{*}{\textbf{\begin{tabular}[c]{@{}c@{}}Standalone RTL \\ implementation \\ with \\ formal property \\ verification\end{tabular}}} &
		\multirow{2}{*}{Llama3.3-70b} &
		No &
		9 &
		3 &
		6 &
		0.654* &
		\multirow{8}{*}{98} &
		22 &
		76 &
		62.13 &
		\multirow{8}{*}{93.28} &
		62.13 \\
		&                               & Yes & 10 & 10 & 0 & 0.046   &  & 26 & 72 & 67.05 &  & 67.05 \\ \cline{2-7} \cline{9-11} \cline{13-13} 
		& \multirow{2}{*}{GPT-4o}       & No  & 12 & 3  & 9 & 1.662*  &  & 22 & 76 & 62.13 &  & 62.13 \\
		&                               & Yes & 11 & 11 & 0 & 0.857   &  & 32 & 66 & 69.34 &  & 69.34 \\ \cline{2-7} \cline{9-11} \cline{13-13} 
		& \multirow{2}{*}{GPT-4.1}      & No  & 14 & 10 & 4 & 0.759*  &  & 22 & 76 & 66.07 &  & 66.07 \\
		&                               & Yes & 11 & 11 & 0 & 0.004   &  & 41 & 57 & 79.34 &  & 79.34 \\ \cline{2-7} \cline{9-11} \cline{13-13} 
		& \multirow{2}{*}{GPT-5}        & No  & 18 & 12 & 6 & 1.109*  &  & 85 & 13 & 90.66 &  & 90.66 \\
		&                               & Yes & 21 & 21 & 0 & 0.092   &  & 94 & 4  & 92.30 &  & 92.30 \\ \hline
	\end{tabular}
	\vspace{1ex}
	\noindent
	\begin{minipage}{\textwidth}
		\vspace{1em}
		\scriptsize 
		Notes: \ac{LLM}-generated properties used a temperature of 1.0.
		Dead code was excluded from coverage in all scenarios.
		Failed assertions generate CEX.
		Formal coverage shows the percentage of "covered and checked" items based on  checker setting with the total cover  items.
		Stimuli coverage shows the percentage of "covered" items with total cover items.
		Checker coverage shows the percentage of "checked" items based on checker setting with the total cover items.\\
		\textit{*Shorter proof times for \ac{LLM} generated properties result from early termination due to \acp{CEX}, not increased verification efficiency.}
	\end{minipage}
\end{table*}

\begin{figure}[t]
	\centering
	\includegraphics[width=\linewidth]{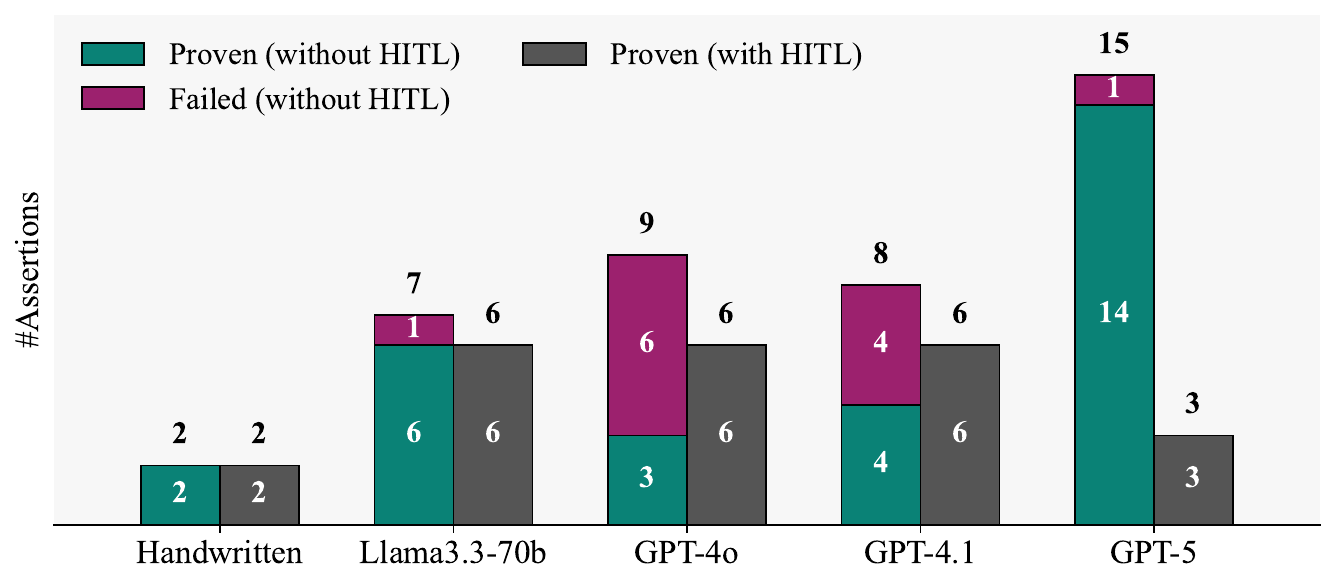}
	\caption{Comparison of handwritten and LLM-based assertion generation for RTL-to-RTL model checker. “With HITL” indicates that counterexamples are fixed during the assertion generation process.}
	\label{fig:assertion_generation_comparison}
\end{figure}

The results in Table~\ref{tab:results} show that \ac{RTL}-to-\ac{RTL} model checking delivers high coverage efficiency. Handwritten assertions achieved 98.42\% formal coverage using only two properties and a proof time of 0.073s. In contrast, \ac{LLM}-generated properties initially required 7–9 assertions to reach similar coverage, with proof times ranging from 1.697s to 7.949s. Although GPT-5 generated up to 15 assertions, it did not improve coverage, revealing redundancy in \ac{LLM}-generated property sets, where multiple properties repeated the same verification intent for behaviors such as exponent alignment. \ac{HITL} refinement significantly improved both correctness and efficiency by eliminating redundant assertions, resolving \acp{CEX} caused by unconstrained signal comparisons, and enforcing architecture-aware conditions for mantissa alignment and rounding behavior. After refinement, concise and valid property sets of 3–6 assertions were produced with no \acp{CEX} across all \acp{LLM}, as shown in Fig.~\ref{fig:assertion_generation_comparison}. Notably, GPT-5 achieved the best efficiency with \ac{HITL}, requiring only three assertions and 1.318s proof time to match the coverage of other \acp{LLM}, as illustrated in Listing~\ref{lst:gpt5_assertions}.

\begin{lstlisting}[
	language=Verilog,
	caption={GPT-5 generated assertions with HITL for model checking},
	label={lst:gpt5_assertions}
	]
	property equal_inputs_outputs_sign_match;
	((impl.s1 == spec.s1) && (impl.s2 == spec.s2) &&
	(impl.e1 == spec.e1) && (impl.e2 == spec.e2) &&
	(impl.m1 == spec.m1) && (impl.m2 == spec.m2))
	|-> (impl.s == spec.s);
	endproperty
	
	property equal_inputs_outputs_exp_match;
	((impl.s1 == spec.s1) && (impl.s2 == spec.s2) &&
	(impl.e1 == spec.e1) && (impl.e2 == spec.e2) &&
	(impl.m1 == spec.m1) && (impl.m2 == spec.m2))
	|-> (impl.e == spec.e);
	endproperty
	
	property equal_inputs_outputs_mant_match;
	((impl.s1 == spec.s1) && (impl.s2 == spec.s2) &&
	(impl.e1 == spec.e1) && (impl.e2 == spec.e2) &&
	(impl.m1 == spec.m1) && (impl.m2 == spec.m2))
	|-> (impl.m == spec.m);
	endproperty
	
	ap_equal_inputs_outputs_sign_match:
	assert property(equal_inputs_outputs_sign_match);
	
	ap_equal_inputs_outputs_exp_match:
	assert property(equal_inputs_outputs_exp_match);
	
	ap_equal_inputs_outputs_mant_match:
	assert property(equal_inputs_outputs_mant_match);
\end{lstlisting}

\begin{figure}[h]
	\centering
	\includegraphics[width=\linewidth]{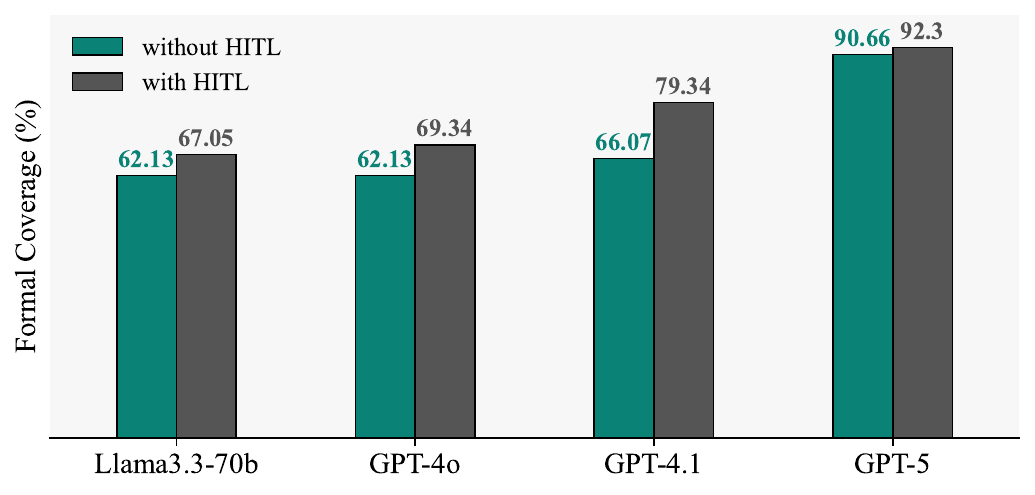}
	\caption{Coverage results without a reference model}
	\label{fig:standalone_coverage_results}
\end{figure}

In the standalone \ac{RTL} verification scenario without a reference model, coverage dropped significantly across all \acp{LLM} and the number of unreachable cover items increased, indicating insufficient design constraints. GPT-5 achieved the highest standalone coverage of 92.30\% but only after generating 21 targeted assertions with \ac{HITL}, as shown in Fig.~\ref{fig:standalone_coverage_results}. This result highlights the difficulty of constraining valid design behavior without micro-architectural guidance. Across all \acp{LLM}, substantial \ac{HITL} refinement was required to correct invalid assumptions, remove properties that produced \acp{CEX}, and add micro-architectural checks specific to floating-point arithmetic, such as sticky-bit propagation, normalization depth, and exponent underflow handling. Taken together, these results show that current \acp{LLM} do not yet adequately capture the reachable state space in the standalone setting, leading to persistent coverage gaps and a high number of unreachable cover items, as illustrated in Table~\ref{tab:results}.

The present evaluation is limited to a single-precision floating-point adder design and therefore does not quantify how coverage or \ac{HITL} effort scale with more complex designs. The proposed approach is expected to extend to related units such as a floating-point subtractor, and its applicability and scalability for more complex \acp{FPU}, such as floating-point multipliers and dividers, will be investigated in future work.

\section{Conclusion} \label{sec:conclusion}
This paper presented a methodology for verifying floating-point arithmetic using direct \ac{RTL}-to-\ac{RTL} model checking, avoiding reliance on high-level C abstractions. Hierarchical decomposition and \ac{CEX}-guided refinement enabled precise fault localization and ensured functional equivalence between the implementation and a golden reference model. Fault injection further validated verification robustness by exposing precision-critical datapath defects.

The study also evaluated agentic AI for property generation. With a reference model, \ac{LLM}-generated properties achieved coverage comparable to handwritten assertions, though they often included overlapping properties that did not contribute new verification intent. Without a reference model, standalone \ac{RTL} verification yielded reduced coverage and increased unreachable logic, requiring substantial \ac{HITL} refinement to restore coverage by constraining valid micro-architectural behavior. These results demonstrate that \ac{RTL}-to-\ac{RTL} model checking supported by AI-generated properties and guided by expert feedback provides a more efficient verification strategy than standalone formal verification for complex arithmetic datapaths. This methodology establishes a foundation for scalable AI-assisted formal verification, with future gains expected through improved extraction of micro-architectural intent to guide \ac{LLM}-based property synthesis and minimize human refinement effort.

\printbibliography

@article{10.1145/307988.307989,
	author = {Kern, Christoph and Greenstreet, Mark R.},
	title = {{Formal verification in hardware design: a survey}},
	journal = {ACM Trans. Des. Autom. Electron. Syst.},
	year = {1999}
}

@online{siemens2024formal,
	author = {Tusinschi, Nicolae},
	title = {{Understanding Formal Verification}},
	journal = {Siemens Verification Horizons Blog},
	year = {2024},
	url = {https://blogs.sw.siemens.com/verificationhorizons/}
}

@misc{bsi_formal_methods,
	author       = {{Federal Office for Information Security (BSI)}},
	title        = {{Formal Methods for Safe and Secure Computer Systems}},
	year         = {2013}
}

@inproceedings{hsieh2024formal,
	author       = {Blaine Hsieh and Stewart Li and Mark Eslinger},
	title        = {{Every Cloud -- Post-Silicon Bug Spurs Formal Verification Adoption}},
	booktitle    = {DVCon},
	year         = {2015}
}

@article{ho2010post_silicon_debug,
	author       = {C. Richard Ho and Michael Theobald and Brannon Batson and J.P. Grossman and Stanley C. Wang and Joseph Gagliardo and Martin M. Deneroff and Ron O. Dror and David E. Shaw},
	title        = {{Post-Silicon Debug Using Formal Verification Waypoints}},
	journal      = {D. E. Shaw Research Publications},
	year         = {2009}
}

@ARTICLE{8766229,
	author={IEEE},
	title={{Standard for Floating-Point Arithmetic}},
	journal={IEEE Std 754},
	year={2019}
	}

@misc{tiemeyer2019cresthardwareformalverification,
	author={Andreas Tiemeyer and Tom Melham and Daniel Kroening and John O'Leary},
	title={{CREST: Hardware Formal Verification with ANSI-C Reference Specifications}}, 
	year={2019},
	archivePrefix={arXiv},
	primaryClass={cs.PL},
	howpublished={arXiv} 
}

@INPROCEEDINGS{8551480,
	author={Qurat-ul-Ain and Hasan, Osman and Saghar, Kashif},
	booktitle={15th HONET-ICT}, 
	title={{Automatic Formal Verification of Digital Components of IoTs Using CBMC}}, 
	year={2018},
	howpublished={IEEE}
}

@INPROCEEDINGS{7308670,
	author={Mukherjee, Rajdeep and Kroening, Daniel and Melham, Tom},
	booktitle={IEEE Computer Society Annual Symposium on VLSI}, 
	title={{Hardware Verification Using Software Analyzers}}, 
	year={2015},
	howpublished={IEEE}
}

@INPROCEEDINGS{9806218,
	author={Abderehman, Mohammed and Rakesh Reddy, Theegala and Karfa, Chandan},
	booktitle={23rd ISQED}, 
	title={{DEEQ: Data-driven End-to-End EQuivalence Checking of High-level Synthesis}}, 
	year={2022},
	howpublished={IEEE}}

@ARTICLE{8759950,
	author={Ludwig, Tobias and Urdahl, Joakim and Stoffel, Dominik and Kunz, Wolfgang},
	journal={IEEE TCAD}, 
	title={{Properties First—Correct-By-Construction RTL Design in System-Level Design Flows}}, 
	year={2020},
	howpublished={IEEE}
}

@INPROCEEDINGS{9489148,
	author={R, Aarthi and C, Aishwarya and U, Akash M and P, Krupasankar and G, Yadukrishnan and P, Anita J},
	booktitle={6th ICCES}, 
	title={{Property Driven Design based Verification for Register Transfer Level Hardware}}, 
	year={2021},
	howpublished={IEEE}
}

@inproceedings{morini2024dpas,
	title = {{Achieving End-to-End Formal Verification of Large Floating-Point Dot Product Accumulate Systolic Units}},
	author = {Morini, Emiliano and Zorn, Bill and Puri, Disha and Eranki, Madhurima and Jampana, Shravya},
	booktitle = {DVCon},
	year = {2024}
}

@inproceedings{pouarz2024sle,
	title = {{Efficient and Exhaustive Floating Point Verification Using Sequential Equivalence Checking}},
	author = {Travis W. Pouarz and Vaibhav Agrawal},
	booktitle = {DVCon},
	year = {2017}
}

@online{pulp-fpu,
	author       = {{PULP Platform}},
	title        = {{pulp-platform/fpu: Floating Point Unit (FPU)}},
	year         = {2018},
	url          = {https://github.com/pulp-platform/fpu}
}

@inproceedings{beers2001applications,
	title={{Applications of hierarchical verification in model checking}},
	author={Beers, Robert and Ghughal, Rajnish and Aagaard, Mark},
	booktitle={{Correct Hardware Design \& Verification Methods}},
	year={2001},
}

@INPROCEEDINGS{11218681,
	author={Gadde, Deepak Narayan and Radhakrishna, Keerthan Kopparam and Viswambharan, Vaisakh Naduvodi and Kumar, Aman and Lettnin, Djones and Kunz, Wolfgang and Simon, Sebastian},
	booktitle={38th SBCCI}, 
	title={{Hey AI, Generate Me a Hardware Code! Agentic AI-based Hardware Design \& Verification}}, 
	year={2025}
}

@misc{cadence_jasper,
	author       = {{Cadence Design Systems}},
	title        = {{Jasper Formal Verification Platform}},
	howpublished = {\url{https://www.cadence.com/}}
}

\end{document}